\documentclass[aps,pra,reprint,superscriptaddress,longbibliography,floatfix]{revtex4-1}

\usepackage{amsmath,braket}
\usepackage{graphicx}
\usepackage{color}
\usepackage{xcolor}
\usepackage[hypertexnames=false,colorlinks,linkcolor={red!50!black},citecolor={blue!50!black},urlcolor={blue!80!black}]{hyperref} 

\usepackage[separate-uncertainty=true,multi-part-units=single]{siunitx}

\definecolor{darkgreen}{rgb}{0.0,0.4,0.0}

\DeclareSIUnit{\mu}{\micro\meter}
\DeclareSIUnit{\unit}{\relax}

\newcommand{\beginsupplement}{%
        \setcounter{table}{0}
        \renewcommand{\thetable}{S\arabic{table}}%
        \setcounter{figure}{0}
        \renewcommand{\thefigure}{S\arabic{figure}}%
     }

\begin{document}

\title{Optical nanoscopy via quantum control\\ \vspace{10PT}
\small{\normalfont{One-sentence-summary:}} \\A scheme for nanoscopic imaging of a coherent quantum mechanical system, a two-level atom, using a far-field optical probe.}

\author{Timo Kaldewey}
\email{timo.kaldewey@unibas.ch}
\affiliation{Department of Physics, University of Basel, Klingelbergstrasse 82, CH-4056 Basel, Switzerland}

\author{Andreas V.\ Kuhlmann}
\altaffiliation[current address: ]{IBM Research-Zurich, S\"{a}umerstrasse 4, 8803 R\"{u}schlikon, Switzerland}
\affiliation{Department of Physics, University of Basel, Klingelbergstrasse 82, CH-4056 Basel, Switzerland}

\author{Sascha R.\ Valentin}
\affiliation{Lehrstuhl f\"{u}r Angewandte Festk\"{o}rperphysik, Ruhr-Universit\"{a}t Bochum, D-44780 Bochum, Germany}

\author{Arne Ludwig}
\affiliation{Lehrstuhl f\"{u}r Angewandte Festk\"{o}rperphysik, Ruhr-Universit\"{a}t Bochum, D-44780 Bochum, Germany}

\author{Andreas D.\ Wieck}
\affiliation{Lehrstuhl f\"{u}r Angewandte Festk\"{o}rperphysik, Ruhr-Universit\"{a}t Bochum, D-44780 Bochum, Germany}

\author{Richard J.\ Warburton}
\affiliation{Department of Physics, University of Basel, Klingelbergstrasse 82, CH-4056 Basel, Switzerland}

\date{\today}

\begin{abstract}
We present a scheme for nanoscopic imaging of a quantum mechanical two-level system using an optical probe in the far-field. Existing super-resolution schemes require more than two-levels and depend on an incoherent response to the lasers. Here, quantum control of the two states proceeds via rapid adiabatic passage. We implement this scheme on an array of semiconductor self-assembled quantum dots. Each quantum dot results in a bright spot in the image with extents down to 30 nm ($\lambda/31$). Rapid adiabatic passage is established as a versatile tool in the super-resolution toolbox.
\end{abstract}


\maketitle 

\section{Introduction}

The diffraction limit prevents a conventional optical microscope from imaging at the nano-scale. Significantly however, diffraction-unlimited imaging of individual molecules is possible by creating an intensity-dependent molecular switch \cite{Betzig2006,Hell2009}. For instance, molecular fluorescence can be turned on with green laser light (absorption of a green photon followed by relaxation), and subsequently turned off with red laser light (stimulated depletion). Fluorescence is only allowed when the power of the red laser is below threshold: the red laser acts as an on-off switch. This switching capability is translated into a microscopy scheme called STED (stimulated emission depletion microscopy) through a ``doughnut" intensity profile of the red beam \cite{Hell1994,Klar2000,Willig2007,Rittweger2009}. Practical considerations and not diffraction per se limit the resolving power of such an imaging scheme. A number of variants on this basic scheme exist \cite{Hell2007,Hell2009,Weisenburger2015}. However, all these schemes use more than two quantum states -- the STED protocol for example uses at least four quantum states (two separate vibronic transitions) -- and all the schemes exploit an incoherent response to the lasers. Missing is a protocol for diffraction-unlimited imaging of a coherent two-level system (TLS).

\begin{figure*}[htbp]{}
	\centering
	\includegraphics[scale=1.0]{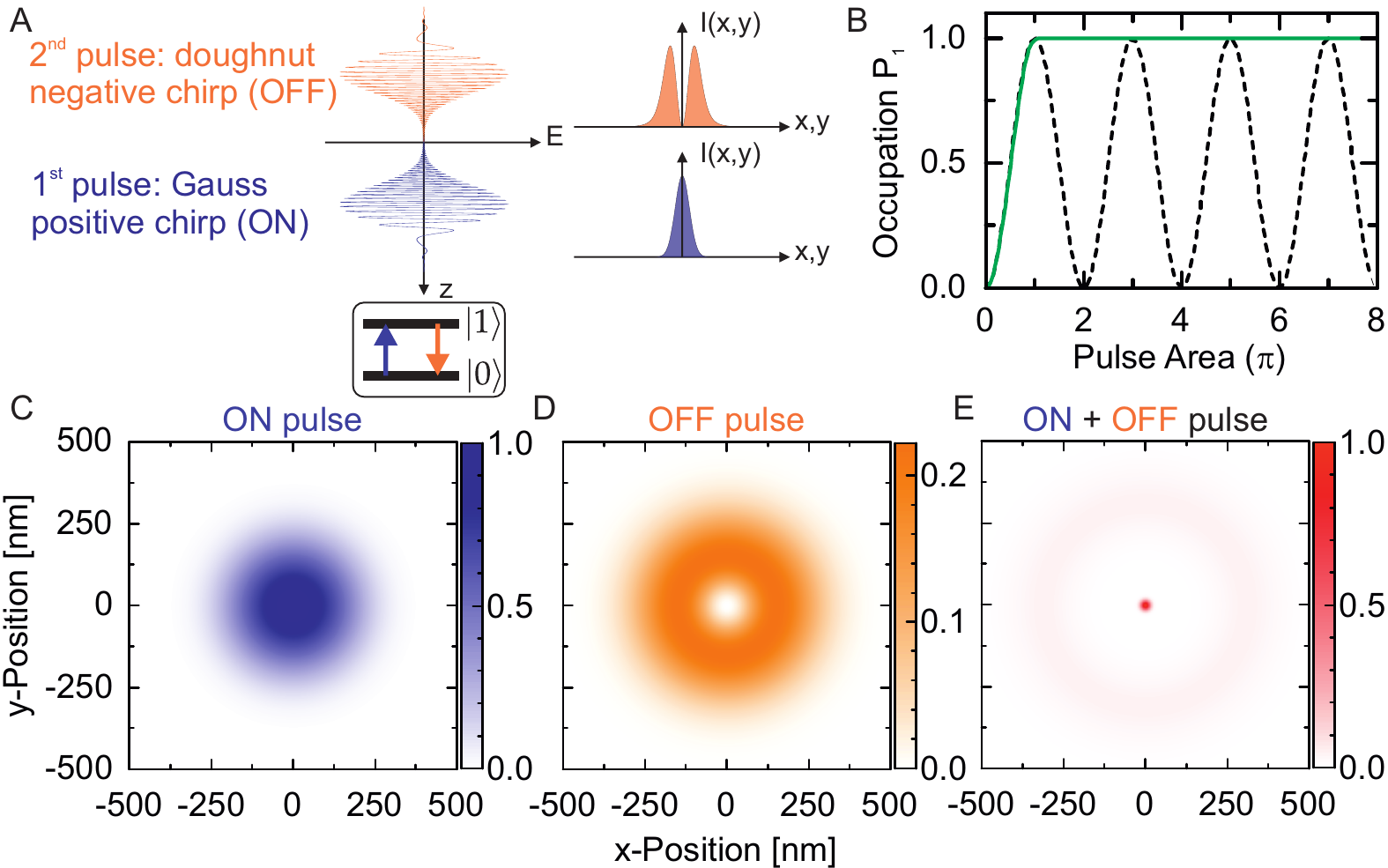}
	\caption{
	\textbf{Concept of nanoscopic imaging of a quantum mechanical two-level system.}
	\textbf{(A)} Temporal waveform and spatial intensity profile of the two optical pulses: excitation (positive chirp, Gaussian profile) and de-excitation (negative chirp, doughnut profile) pulse. \textbf{(B)} Response of an ideal two-level system to a single laser pulse. The system is initially in the ground state $\ket{0}$; plotted is the occupation of the excited state $\ket{1}$, $P_{1}$, as a function of pulse area. A resonant, unchirped pulse drives a Rabi oscillation (dashed black line). A chirped pulse (green solid line) transfers the system from state $\ket{0}$ to $\ket{1}$ for pulse areas above $\pi$ by means of rapid adiabatic passage. 
	\textbf{(C) - (E)} Simulation of the imaging experiment: $P_{1}$ as a function of sample $(x,y)$-position. (C) Gauss-pulse only with $I_{G}^0=1 I_T$; (D) doughnut-pulse only with $I_{D}^0=1 I_T$; (E) Gauss-pulse $I_{G}^0=3 I_T$ followed by doughnut-pulse $I_{D}^0=205 I_T$ resulting in a 15 times smaller spot size. Our model for the simulation is described in the supplementary material \cite{MaM}. Here we used the following parameters: wavelength \SI{940}{\nano\meter}, refractive index solid immersion lens $n_{\mathrm{SIL}}=2.13$, beam diameter before the objective $\Delta X_{I,\mathrm{FWHM}}=\SI{2.0}{\milli\meter}$, focal length $f=\SI{3.7}{\milli\meter}$, detection efficiency $\beta=1$, chirp $\alpha_{\mathrm{G}}=-\alpha_{\mathrm{D}}=\SI{3.24}{\pico\second^{-2}}$.
	}
  \label{fig:fig1_concept}
\end{figure*}

The coherent response of a TLS to a pulse of resonant laser light is a Rabi oscillation, a rotation of the quantum state around the Bloch sphere. For instance, the system is driven from its ground state $\ket{0}$ to the excited state $\ket{1}$ and then back to $\ket{0}$ and so on, Fig.\ \ref{fig:fig1_concept}(B). In terms of fluorescence, Rabi oscillations represent an off-on-off-on\ldots behavior as a function of pulse area. If the TLS is illuminated with a high intensity pulse with Gaussian spatial profile then a series of rings in the optical response results \cite{Gerhardt2010}. In other words, Rabi oscillations do not represent a simple on-off switch and are not well suited to imaging. We propose instead that rapid adiabatic passage (RAP) is an excellent technique for imaging a TLS. In RAP, the laser frequency is swept through the resonance of the TLS during the pulse. RAP transfers the population from one state to the other but without the oscillations at large pulse energies, Fig.\ \ref{fig:fig1_concept}(B). Applied to an ideal TLS initially in state $\ket{0}$, a weak RAP pulse leaves the system in state $\ket{0}$ whereas a strong RAP pulse transfers the system to state $\ket{1}$; and vice versa for the system initially in state $\ket{1}$. The clear threshold of RAP on a TLS represents an ideal on-off switch with immediate applications in imaging.

Our scheme is shown in Fig.\ \ref{fig:fig1_concept}. The TLS is initially in the ground state $\ket{0}$. A pulse with positive chirp and above-threshold pulse area is applied. This inverts the TLS provided it is located somewhere within the diffraction-limited spot: it turns the system on. Subsequently, a pulse with negative chirp and above-threshold pulse area is applied. This pulse inverts the system a second time, leading to re-occupation of state $\ket{0}$: it turns the system off. However, when the first pulse has a Gaussian intensity distribution (Fig.\ \ref{fig:fig1_concept}(C)) and the second pulse a doughnut intensity distribution (Fig.\ \ref{fig:fig1_concept}(D)), the second pulse is inactive at the center of the doughnut. Under these conditions, the system is left in the upper ``on" state $\ket{1}$ only when it is located close to the center of the doughnut. The system is then left to decay by spontaneous emission and the photon is detected. In this way, a fluorescence bright spot results at the center of the doughnut. The resolution in the image is determined by the spatial location at which the doughnut intensity crosses the RAP threshold, a ``physics-based diffraction-unlimited" resolution \cite{Hell2009}. This imaging scheme can be described analytically in a very simple way (Gaussian optics for the two beams, the Landau-Zener formalism \cite{Shevchenko2010} for RAP): Fig.\ \ref{fig:fig1_concept}(E) shows how a bright spot emerges at the doughnut center. The full model is described in the supplementary material \cite{MaM}.

Establishing an imaging protocol based on RAP strikes us as important. First, it opens up a way to image on the nano-scale coherent quantum optical systems such as cold atoms and trapped ions, also their solid-state counterparts, semiconductor quantum dots and color centers. Second, in a multi-level system, RAP avoids the excitation of molecular vibrations (phonons in a solid-state context). In STED microscopy, close to an eV of energy is dumped into the molecule per cycle, leading to blinking and bleaching. This heating can be avoided in RAP (by choosing positive chirp for the $\ket{0} \rightarrow \ket{1}$ process, negative chirp for the $\ket{1} \rightarrow \ket{0}$ process \cite{Luker2012,Wei2014}). Third, the chirped laser pulses are spectrally broadband yet the fluorescence is narrowband. This combination allows an inhomogeneous distribution of emitters (distribution in space or in time) to be addressed with the same laser pulses. Any spectral fingerprint in the emitters is retained by spectrally resolving the fluorescence, for instance with a spectrometer and array detector. Finally, our test system, a semiconductor quantum dot, is an extended emitter, with lateral extent $\sim 5$ nm. Although a quantum dot mimics a real atom it has a much larger size. Our scheme opens the perspective of imaging an electron wave function with a far-field, optical microscope.

\section{Results}

\begin{figure*}[htbp]{}
	\centering
	\includegraphics[scale=1.0]{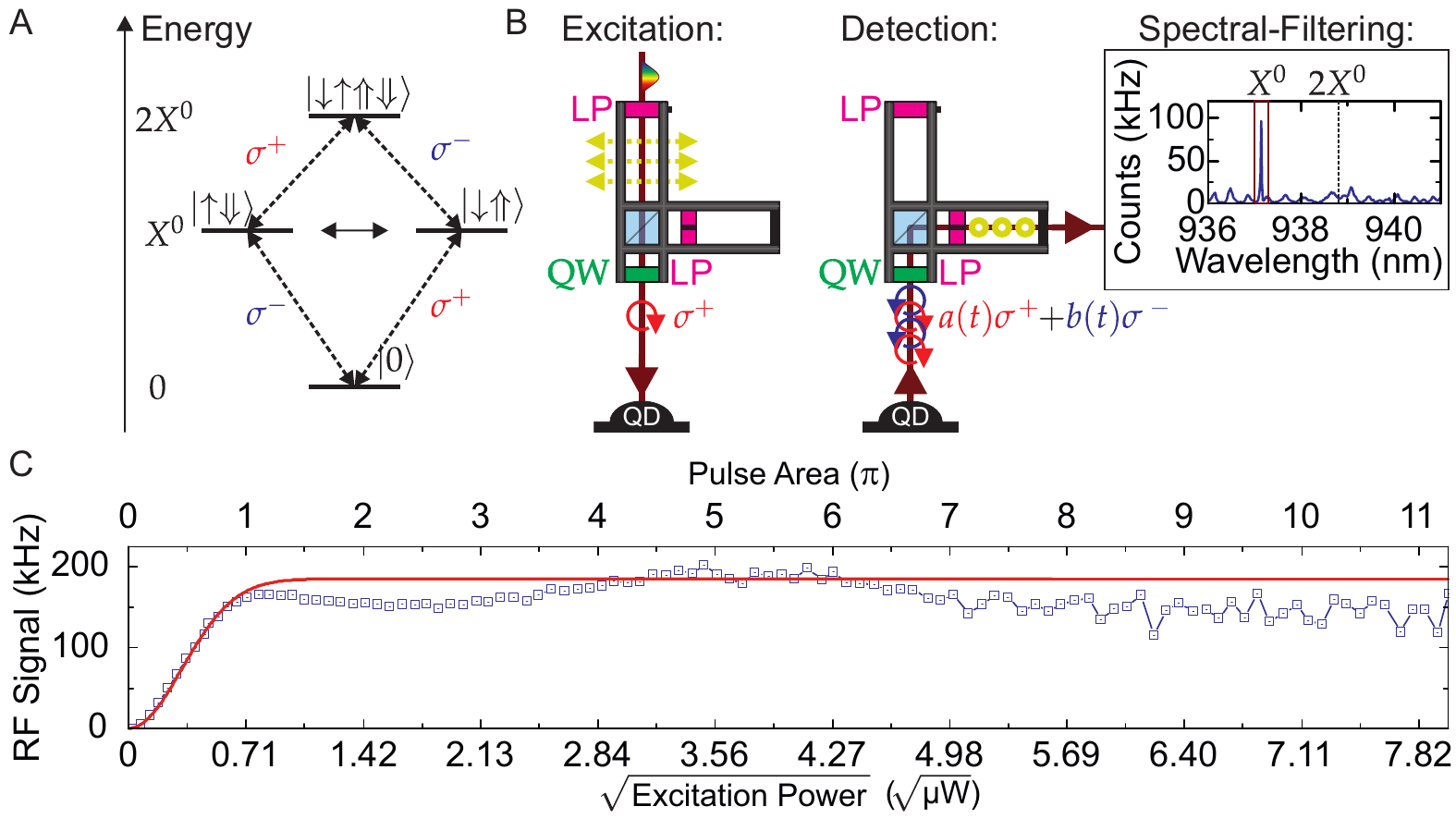}
	\caption{\textbf{Rapid adiabatic passage on a single self-assembled quantum dot.} 
	\textbf{(A)} Energy level scheme of the quantum dot (QD): $\ket{0}$ represents the empty QD; $\ket{1}$ the spin-up exciton, $\ket{1} \equiv \ket{\downarrow\Uparrow}$. ($\uparrow$ ($\downarrow$) is a spin-up (spin-down) electron, $\Uparrow$ ($\Downarrow$) a spin-up (spin-down) hole) The transition $\ket{0} \leftrightarrow \ket{\downarrow\Uparrow}$ is driven with right-handed circularly polarised light ($\sigma^{+}$). The two exciton states, $\ket{\downarrow\Uparrow}$ and $\ket{\uparrow\Downarrow}$, are coupled via the fine structure which leads to a quantum beat $\ket{\downarrow\Uparrow} \leftrightarrow \ket{\uparrow\Downarrow}$. On creating $\ket{\downarrow\Uparrow}$ with a $\sigma^{+}$ polarised pulse results hence in both $\sigma^{+}$- and $\sigma^{-}$ polarised emission. The biexciton state, 2X$^{0}$, exists at higher energies but is not populated here. 
	\textbf{(B)} Polarisation control in the dark-field microscope: the QD is excited with $\sigma^{+}$ polarised light; $\sigma^{-}$ is detected. LP refers to a linear polariser, QW to a quarter-wave plate. The detected signal is spectrally filtered. This increases the signal to background ratio. 
	\textbf{(C)} Resonance fluorescence versus pulse area (and square root of averaged excitation power $P_{\mathrm{avg}}$) on a single, empty quantum dot. Plotted is the detected emission from the $\ket{X^{0}} \rightarrow \ket{0}$ transition following circularly polarised excitation pulses (blue squares). The originally transform limited pulses ($\tau=\SI{80}{\femto\second}$) were positively chirped ($\phi_2=\SI{0.33}{\pico\second^2}$) to a pulse duration of \SI{4}{\pico\second}. The data are fitted to the Landau-Zener result (red line) with $A(1-\exp{(-c^2 \tau \sqrt{\tau^4+\phi_{2}^2}/\phi_{2}*P_{\mathrm{avg}}})$ and reveal $c\sqrt{\tau}=\SI{4.4\pm0.2}{\micro\watt^{-1/2}}$ in agreement with independently estimated values from Rabi oscillations (\SI{4.4\pm0.4}{\micro\watt^{-1/2}}). The parameter c includes the dipole transition moment and when multiplied with $\sqrt{\tau P_{\mathrm{avg}}}$ gives the pulse area.
	}
  \label{fig:fig2_RAP_QD}
\end{figure*}

We use a single quantum dot to implement the RAP-based imaging scheme. At low temperature with resonant optical driving, an InGaAs quantum dot embedded in GaAs mimics closely a two-level atom with radiative lifetime \SI{800}{\pico\second} and emission wavelength \SI{950}{\nano\meter} \cite{Muller2007,Nguyen2011,Matthiesen2012,Kuhlmann2013_RSI}. Here, state $\ket{0}$ is the crystal ground state (empty quantum dot), state $\ket{1}$ is an electron-hole pair (the neutral exciton, X$^{0}$), the result of promoting the highest energy valence electron across the fundamental gap to the lowest energy conduction state, Fig.\ \ref{fig:fig2_RAP_QD}(A). There are two optically-allowed excitons with spin $\pm 1$ which can be created with $\sigma^{\pm}$ polarisation ($\sigma$ represents circular polarisation), Fig.\ \ref{fig:fig2_RAP_QD}(A). We detect the spontaneous emission (technically, the ``resonance fluorescence") as the exciton decays to the ground state. We reject back-reflected laser light with a polarisation-based dark-field concept \cite{Vamivakas2009,Ylmaz2010,Kuhlmann2013_RSI,Kuhlmann2013_NatPhys}: we excite with $\sigma^{+}$ and detect with $\sigma^{-}$, Fig.\ \ref{fig:fig2_RAP_QD}(B). (We note that the $\pm 1$ states are coupled together by some symmetry breaking leading to a quantum beat \cite{Tartakovskii2004} between $\ket{+1}$ and $\ket{-1}$ with period $\sim 100$ ps, Fig.\ \ref{fig:fig2_RAP_QD}(A). This coupling allows us to detect the decay of excitons with both spins.) The chirped laser pulses are created from transform-limited pulses from a mode-locked laser by introducing wavelength-dependent phase shifts. Unusually for quantum dot experiments \cite{Simon2011,Wu2011,Wei2014}, we use the full bandwidth of a 100 fs laser pulse: this allows us to address almost all the quantum dots in the sample with the same laser pulse despite the large inhomogeneous broadening. Fig.\ \ref{fig:fig2_RAP_QD}(C) plots the resonance fluorescence as a function of pulse area on a single quantum dot following excitation with a single laser pulse. The signal rises initially and is then roughly constant above a pulse area of $\pi$. The signal follows closely the Landau-Zener result for a TLS. Only difficulties in rejecting the back-reflected laser pulse at large laser power limit the maximum pulse area to $11 \pi$.

\begin{figure*}[htbp]{}
	\centering
	\includegraphics[scale=1.0]{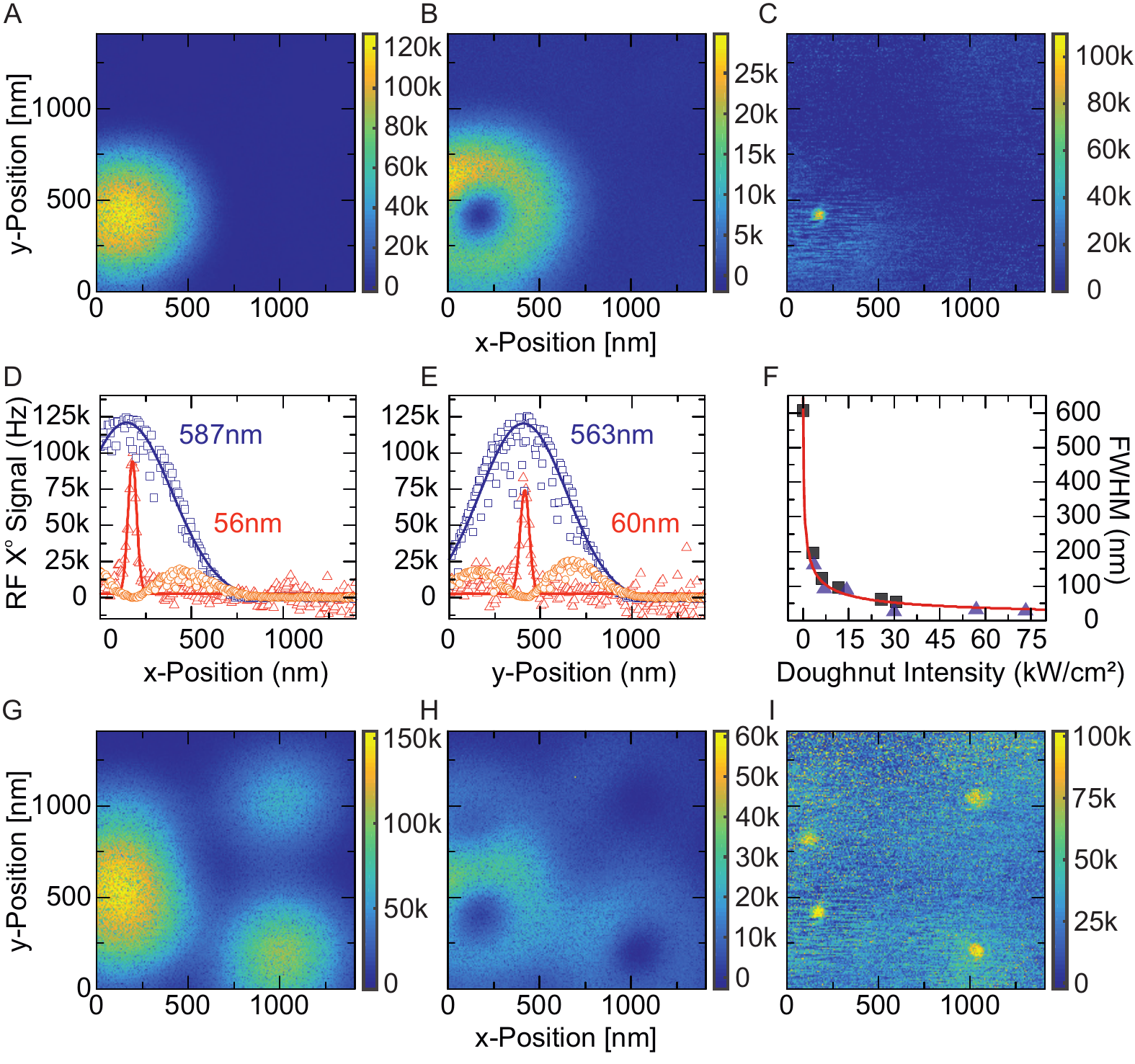}
	\caption{\textbf{Imaging an ensemble of quantum dots (QDs) with the RAP-based protocol.} 
	\textbf{(A)-(C)} Resonance fluorescence as a function of sample position. The excitation intensities are stated in units of the threshold intensity $I_T=\SI[per-mode=symbol]{0.28}{\kilo\watt\per\centi\meter\squared}$. (A) Gauss-pulse only, intensity $I_{G}^0=1.2 I_T$; (B) doughnut-pulse only, intensity $I_{D}^0=1.2 I_T$; (C) Gauss-pulse $I_{G}^0=2.8 I_T$ followed by doughnut-pulse $I_{D}^0=205 I_T=\SI[per-mode=symbol]{57}{\kilo\watt\per\centi\meter\squared}$. Shown is the signal in a spectral window (wavelength bandwidth 0.05 nm) around the X$^{0}$ emission of one single QD. 
	\textbf{(D)} and \textbf{(E)} show $x$ and $y$ line-cuts through the data in (A)-(C) in blue, orange and red, respectively. Solid lines are Gaussian fits to the blue and red data points. \textbf{(F)} Extracted full-width-at-half-maximum (FWHM) of the central image of a single QD as a function of the doughnut-pulse intensity $I_{D}$. Thereby, $I_{G} = 1.2 I_{T}$ blue triangles, and $I_{G} = 2.8 I_{T}$ black squares. The smallest FWHM achieved here is \SI{30}{\nano\meter}. \textbf{(G)-(I)} Images, as (A)-(C), but adding the signal from four integration windows (each with a width in wavelength of \SI{0.05}{\nano\meter}) demonstrating the multiplexing capability of the imaging scheme. The color scale in (I) is slightly over-saturated to increase the contrast for less bright QDs.
	}
  \label{fig:fig3_STED_results}
\end{figure*}

The RAP-based imaging proceeds by implementing the two-pulse scheme, creating an image by sample scanning. We consider initially imaging a single quantum dot by collecting the resonance fluorescence only at its X$^{0}$ emission wavelength, Fig.\ \ref{fig:fig3_STED_results}(A)-(C). The Gauss-pulse alone (pulse area $ \pi$) gives a Gaussian spatial response with mean full-width-at-half-maximum (FWHM) \SI{575}{\nano\meter}, slightly larger than the confocal diffraction-limited spot size on account of the non-linear response of RAP. Similarly, the response to the doughnut-pulse alone gives a doughnut-profile, with a width, as for the Gauss-pulse, determined largely by diffraction. Applying both pulses sequentially breaks the diffraction limit: a bright spot emerges with FWHM \SI{56}{\nano\meter} in this particular experiment. The signal and the signal-to-background ratio at the central bright spot are almost the same as for the Gauss-pulse alone. This is a significant point: RAP-imaging simply concentrates the available signal to a smaller region in the image. 

The FWHM of the central image decreases with increasing doughnut-pulse intensity, Fig.\ \ref{fig:fig3_STED_results}(F). This is the signature of ``physics-based diffraction-unlimited" performance: it is the control of the on-off switching which determines the resolution, not the diffraction-limited focusing of the laser beams. The FWHM $\Delta x$ follows the same functional form as for STED microscopy \cite{Harke2008}, $\Delta x(I_{D})= \Delta x^{o}/\sqrt{1+I_{D}/I_{T}}$ where $\Delta x^{o}$ is the conventional diffraction limit, $I_{D}$ is the dougnut intensity, and $I_{T}$ the off-on threshold intensity \cite{MaM}. Here, we can anticipate $\Delta x(I_{D})$ based only on the characterization of the microscope performance (resonance fluorescence from Gauss beam at intensities well below saturation) and RAP characterization on the quantum dot. The result describes the experimentally determined FWHM extremely well, Fig.\ \ref{fig:fig3_STED_results}(F).

A key application of super-resolution microscopy is to garner an image with detail which is obscured in a diffraction-limited microscopy. We demonstrate this by imaging a region of the sample containing a number of quantum dots separated laterally by distances smaller than or comparable to the diffraction limit of the microscope. An image with the Gauss-pulse alone, equivalently the doughnut-pulse alone, shows structure but it is difficult to determine the number and location of individual emitters, Fig.\ \ref{fig:fig3_STED_results}(G),(H). With the RAP-based imaging, the number and location of the point-like emitters is clearly visible, Fig.\ \ref{fig:fig3_STED_results}(I). The RAP-based imaging in the present experiment is limited only by the technical difficulty of distinguishing the resonance fluorescence signal from the reflected laser light at the highest doughnut-pulse peak areas.

\section{Discussion}
Working with a two-level emitter depends on distinguishing the resonance fluorescence of the emitter from the laser light used to create it. Unlike fluorescence-based imaging there is a spectral overlap between signal and source. The techniques used here (crossed polarisation and spectral analysis) can be supplemented with more effective spectral filtering for individual emitters or a collection of close-to-identical emitters (the source is broadband, resonance fluorescence from an individual emitter is narrowband). Time-gating the detector could also be used to suppress the background laser signal: in this experiment the pulses are gone in $\sim 50$ ps yet the radiative lifetime is much larger, $\sim 800$ ps. With these improvements we believe that the technique can become a versatile one.

\section{Materials and Methods}
The InGaAs quantum dots are embedded in a GaAs n-i-p diode [25] and emit around 950 nm wavelength. The heterostructure is grown by molecular beam epitaxy. The collection efficiency is increased by incorporating a semiconductor Bragg mirror below the quantum dots in the heterostructure growth and by placing a solid immersion lens on the sample surface. Resonance fluorescence is distinguished from back-scattered laser light by a polarisation-based dark-field microscope \cite{Kuhlmann2013_RSI,Kuhlmann2013_NatPhys} and by spectrally-sensitive detection. The laser is a mode-locked Ti:sapphire laser and produces close-to-transform-limited pulses with a duration of \SI{\sim{}100}{\femto\second}. The laser output is split into two beams, and two separate pulse-shapers introduce a controlled amount of chirp into the beams. One beam is delayed with respect to the other and then each beam is coupled into an optical fiber. The two optical fibers transport the beams to the two input ports of the microscope. Following collimation in the microscope, one of the beams passes through a $2\pi$ vortex phase plate. The two beams are then combined at a beam-splitter. The microscope objective, sample and nano-positioners are held at 4.2 K in a helium bath cryostat. On focussing, the beam with the phase-manipulated wavefront acquires a doughnut intensity profile. The doughnut maximum:minimum intensity ratio $>1 000:1$. The microscope objective has numerical aperture 0.68 and operates close to the diffraction limit. Images are recorded by sample scanning.

\begin{acknowledgments}
We acknowledge financial support from EU FP7 ITN S$^{3}$NANO, NCCR QSIT and SNF project 200020\_156637. SRV, AL and ADW acknowledge gratefully support from BMBF Q.com-H 16KIS0109.
\end{acknowledgments}

\bibliography{coherent_RAP_STED_TK_05_ref}

\newpage
\beginsupplement
\section{Supplementary Materials}

%

\subsection{The semiconductor sample}
\subsubsection{The heterostructure}
\begin{figure*}[thbp]{}
	\centering
	\includegraphics[scale=1.0]{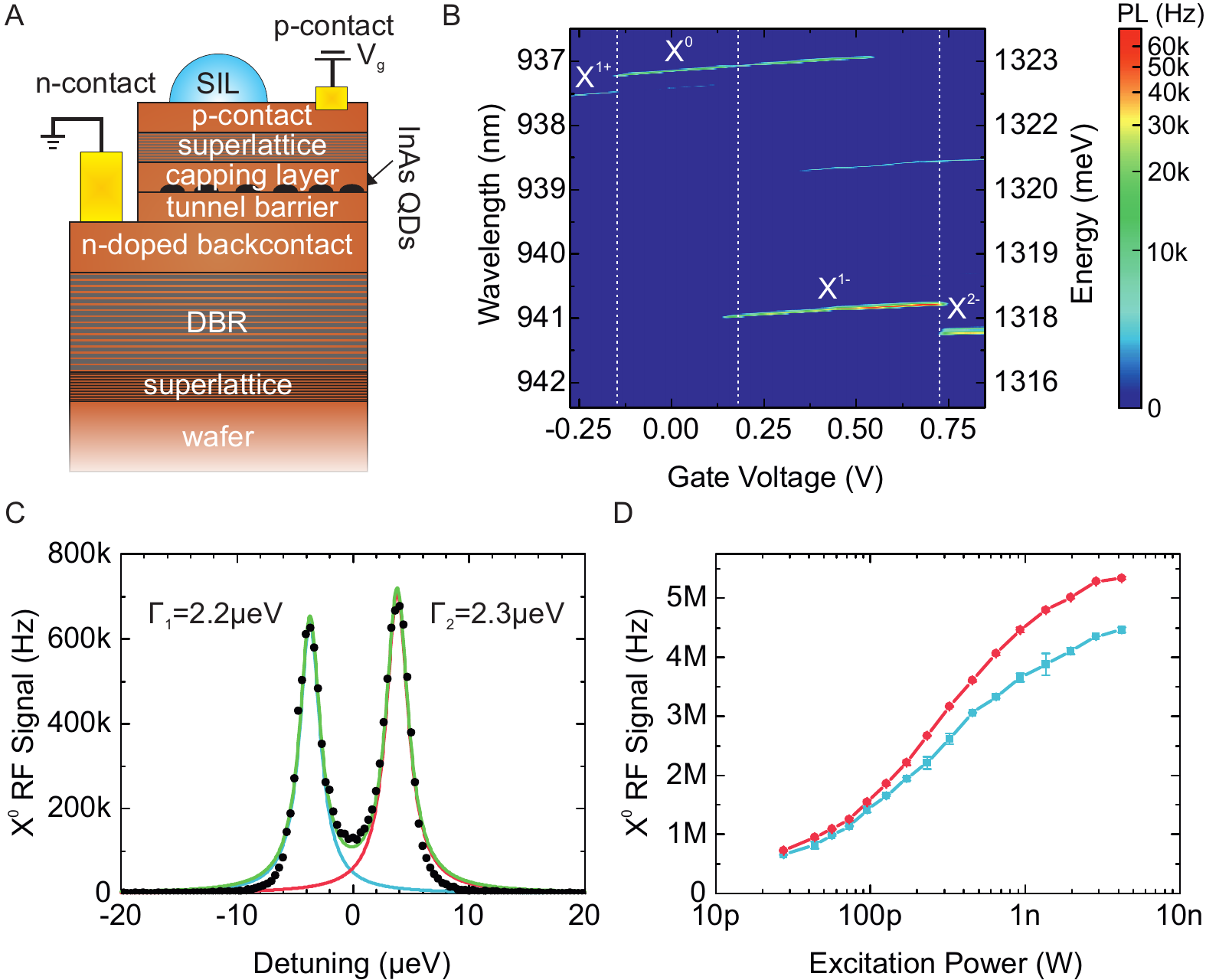}
  \caption{\textbf{Resonance fluorescence on a single quantum dot (QD): CW excitation.} 
	\textbf{(A)} The heterostructure: InGaAs QDs in an n-i-p diode with weak cavity structure. In the growth direction: GaAs wafer, GaAs/AlAs superlattice (18 periods), distributed Bragg reflector (DBR) consisting of 16 GaAs/AlAs pairs, Si doped GaAs back contact, GaAs tunnel barrier, InGaAs QDs, capping layer, blocking barrier consisting of AlAs/GaAs superlattice, C-doped GaAs as top contact. A ZrO$_{2}$ solid immersion lens (SIL) is placed on the sample surface to increase the photon extraction efficiency.
	\textbf{(B)} Photoluminescence of a single QD, QD01, as a function of applied gate voltage and emission wavelength. The QD was excited non-resonantly at \SI{830}{\nano\meter} with a power of \SI{17}{\micro\watt}.
	\textbf{(C)} Resonance fluorescence spectrum of QD01 X$^{0}$ with resonant, circularly polarised excitation of \SI{27}{\pico\watt}. The data (black circles) were fitted with a sum of two Lorentzian functions (green, individual functions shown in red and blue). The individual linewidths are \SI{2.2}{\micro\electronvolt} and \SI{2.3}{\micro\electronvolt}. The photons were detected with a silicon single photon avalanche diode with an integration time of \SI{5}{\milli\second}. The laser frequency was \SI{319.77862}{\tera\hertz} (wavelength \SI{937.530}{\nano\meter}). 
	\textbf{(D)} Fitted amplitudes of resonance fluorescence as a function of excitation power for the two X$^{0}$ transitions.
	}
	\label{fig:sample}
\end{figure*}

The InGaAs quantum dots (QDs) are embedded in a GaAs n-i-p diode grown by molecular beam epitaxy \cite{Prechtel2016}. The layer structure is shown in Fig.\ \ref{fig:sample}(A). Initially, a superlattice (18 periods of a \SI{2}{\nano\meter} GaAs, \SI{2}{\nano\meter} AlAs unit) is grown on top of a GaAs wafer in order to smooth the wafer surface. Following this, a distributed Bragg reflector (DBR) consisting of 16 periods of \SI{68.6}{\nano\meter} GaAs and \SI{81.4}{\nano\meter} AlAs is grown in order to increase the photon collection efficiency from the top side. The back contact is provided by a layer of n-doped GaAs. An intrinsic region, \SI{30}{\nano\meter} of GaAs, acts as tunnel barrier for the InGaAs QDs. The QDs are capped by \SI{153}{\nano\meter} GaAs. A subsequent barrier (a superlattice consisting of 46 periods of a \SI{3}{\nano\meter} AlAs, \SI{1}{\nano\meter} GaAs unit) hinders current flow in the strong vertical electric field. The penultimate layer is p-doped GaAs forming the top contact of the diode. Compared to a Schottky diode structure the absorption of photons in the top gate is reduced. The final layer, \SI{44}{\nano\meter} undoped GaAs, places the p-doped layer around a node-position of the standing electromagnetic wave. The n- and p-doped layers are contacted independently. Selective etching of the capping allows contacting of the buried p-layer. Access to the n-layer is provided by wet etching of a mesa structure around the solid immersion lens (SIL). The n-contact is grounded; a voltage $V_g$ is applied to the p-contact.

\subsubsection{Quantum dot charging}
The sample is probed at liquid helium temperature, \SI{4.2}{\kelvin}, in a bath cryostat. At these temperatures, there is a pronounced Coulomb blockade: the number of electrons in the quantum dot changes step-wise as a function of $V_{g}$ \cite{Warburton2000,Dalgarno2008,Warburton2013}. Fig.\ \ref{fig:sample}(B) shows the photoluminescence (PL) of a single QD as function of $V_{g}$. There are plateaus with emission from the neutral exciton, X$^{0}$, from the negatively charged trion (two-electron, one-hole complex), X$^{1-}$, and from the X$^{2-}$. There is also a much weaker PL signal from the positively charged trion, X$^{1+}$, at the most negative $V_g$. The X$^{0}$ is used in the imaging experiments. We note that the $V_{g}$-overlap of the X$^{0}$-X$^{1-}$ plateau is not a feature once the QD is driven with resonant excitation \cite{Kurzmann2016}.

\subsubsection{Resonance fluorescence detection: CW excitation}
Resonance fluorescence (RF), the scattering induced by resonant excitation, is shown in Fig.\ \ref{fig:sample}(C) on the X$^{0}$ transition. The laser excitation and the detection are orthogonally polarised in order to suppress reflected laser light in the detection channel: this is a polarisation-based dark-field technique \cite{Kuhlmann2013_NatPhys,Kuhlmann2013_RSI}. With our RF multi-purpose microscope, described in sec.\ \ref{ssec:microscope}, CW excitation, and a zirconia SIL on the sample surface, we reach a driving laser extinction ratio of \num{e7}:1 and a collection efficiency of approximately $5-10$\SI{}{\percent}. The net result is that the signal:background ratio is 250:1, e.g.\ Fig.\ \ref{fig:sample}(C). The X$^{0}$ transition is split into two, Fig.\ \ref{fig:sample}(C), as a consequence of the fine structure splitting, FSS (see Fig.\ 1 of the main article). The linewidths are around 2 $\mu$eV, approximately 2.5 times larger than the transform limit \cite{Kuhlmann2013_NatPhys}. Above saturation, the RF is more than \SI{5}{\mega\hertz} corresponding to a single photon flux of \SI{25}{\mega\hertz}. The detector, a silicon single photon avalanche diode (SPAD), has a quantum efficiency of 20\SI{}{\percent} at a wavelength of \SI{950}{\nano\meter}.

\subsection{The experiment}
\begin{figure*}[thbp]{}
	\centering
	\includegraphics[scale=1.0]{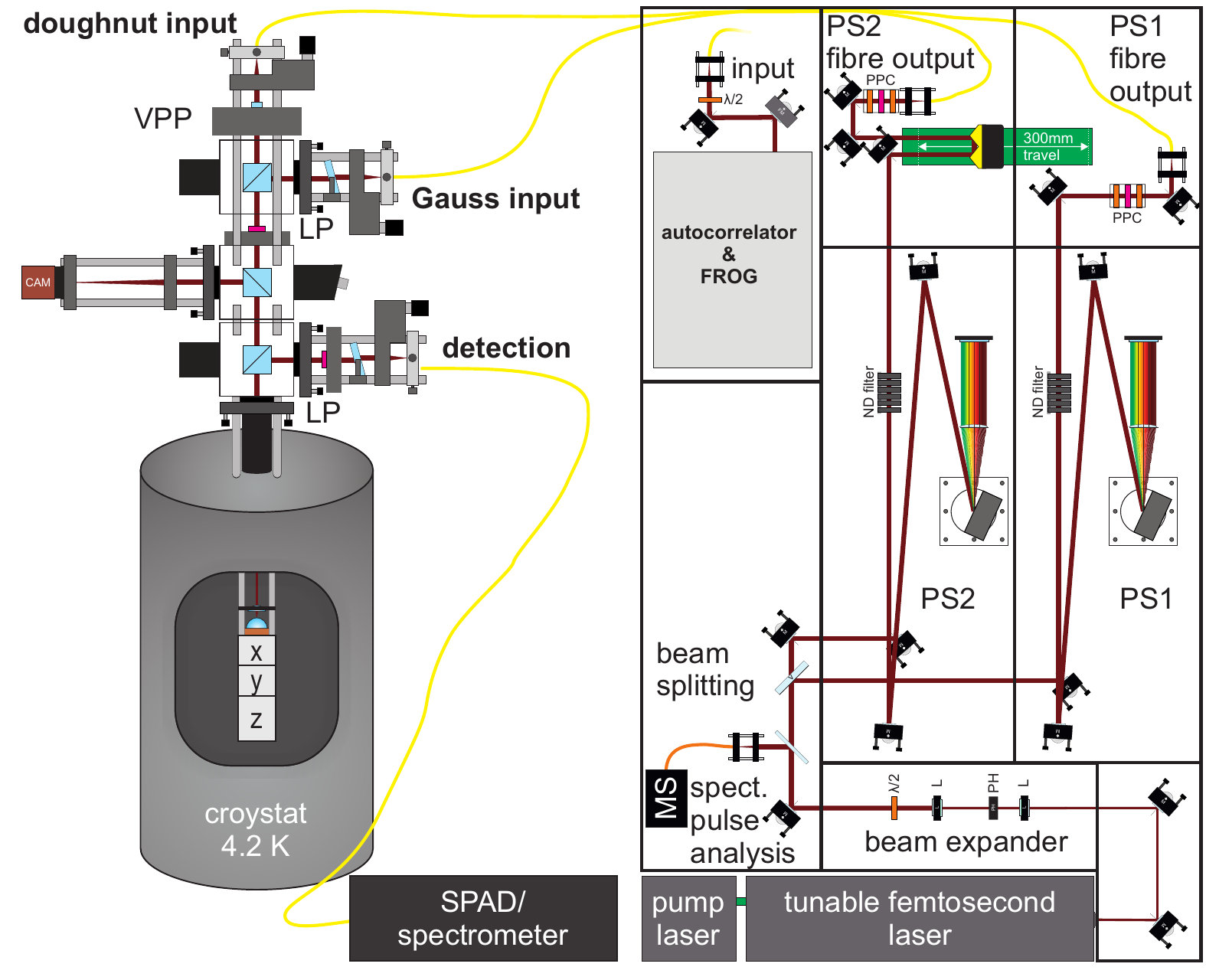}
  \caption{{\bf Scheme of the whole experiment.} L lens, $\lambda/2$ half-wave plate, PH pin-hole, MS mini-spectrometer, PS pulse shaper, ND neutral density filter, PPC power-polarisation-control, FROG frequency-resolved-optical-gating, VPP vortex phase-plate, LP linear polariser, SPAD single photon avalanche diode. The autocorrelator-FROG combination is used to characterise the pulses before the microscope.}
  \label{fig:exp_scheme}
\end{figure*}

The imaging experiment consists of a laser system and a cryogenic optical microscope, Fig.\ \ref{fig:exp_scheme}. The laser is a tunable, mode-locked Ti:sapphire which creates close-to-transform-limited pulses with a pulse duration of \SI{\sim100}{\femto\second}. The pulse train is expanded with a beam expander and split into two. The individual pulse trains are directed into compact pulse shaper units (PS1, PS2) which introduce chirp by introducing a wavelength dependent phase shift. Subsequently, one pulse is delayed with respect to the other in a \SI{300}{\milli\meter} delay line, leading to a delay of up to \SI{2}{\nano\second}, and then power and polarisation are controlled (neutral density filters for coarse power control; a half-wave plate, polariser, half-wave plate combination for fine power control and control of the linear polarisation axis). Finally, the two pulse trains are coupled into separate single-mode optical fibers. 

The microscope consists of the ``head" at room temperature and an objective lens plus nano-positioners at low temperature, Fig.\ \ref{fig:exp_scheme}. The nano-positioners allow the sample to be positioned at the focus of the microscope, and images to be recorded by sample scanning. The head converts one pulse train into a collimated beam with a Gaussian intensity profile, the other into a collimated beam with a Gaussian intensity profile and azimuthal phase dependence. The two pulse trains are focused by the objective onto the sample with close to diffraction-limited performance, resulting in a Gauss-shaped focus and a doughnut-shaped focus. Resonance fluorescence is collected by the same objective and coupled in the microscope head into a third fiber which transports the signal to the detector, either a spectrometer-CCD camera system or a SPAD. 

\subsubsection{Creation and characterization of chirped pulses}
\begin{figure*}[htbp]{}
	\centering
	\includegraphics[scale=1.0]{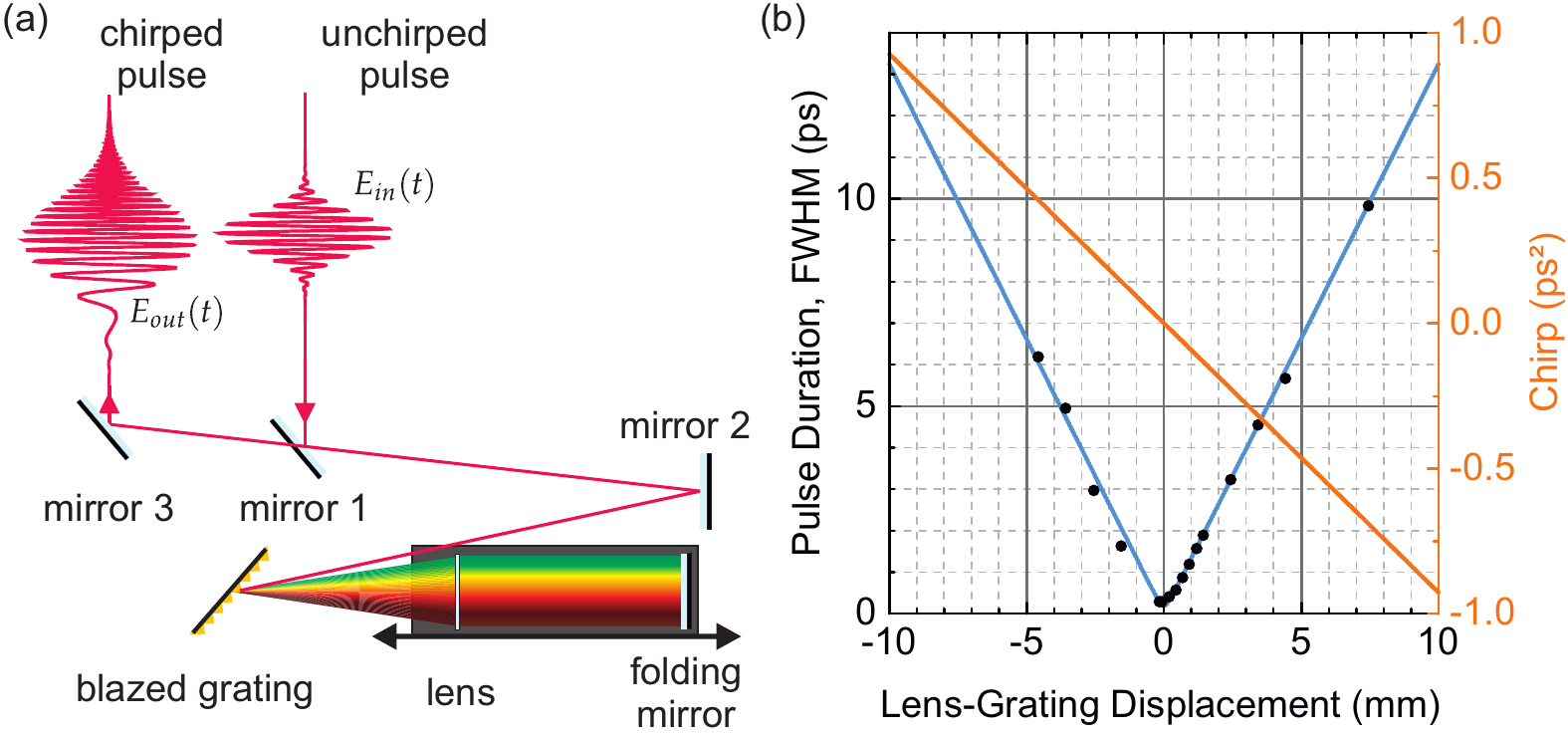}
	\caption{\textbf{Compact pulse shaper: concept and performance.}
	\textbf{(A)} The incident beam is directed with mirrors 1 and 2 onto a blazed grating (1,800 lines/\SI{}{\milli\meter}), diffracted and then focused by a cylindrical lens (focal length 150 mm) onto a mirror placed in the focal plane of the lens. The lens and the folding mirror are mounted on a moving platform. A displacement (up to $\pm10$ \SI{}{\milli\meter}) between grating and lens--mirror assembly introduces chirp. Note that the beam is reflected slightly upwards by the mirror in the pulse shaper such that it overshoots mirror 1 on the return path. 
	\textbf{(B)} Pulse duration $\Delta t_{\mathrm{FWHM}}^I$ as a function of the lens--grating displacement. The pulse duration was measured with a FROG autocorrelator. The measured data points (black circles) are fitted (blue solid line) using the result for quadratic phase, eq.\ \ref{eq:tau}. The sweep rate $\alpha$ calculated from the pulse duration is shown on the right axis.}
	\label{fig:pulse_shaper}
\end{figure*}
A controlled chirp is introduced into the laser pulses using a compact, folded $4f$-pulse shaper \cite{Weiner1992}, Fig.\ \ref{fig:pulse_shaper}(B). Our scheme retains all the frequency components of the femto-second laser. Introducing chirp inevitably results in an increased pulse duration.

An unchirped pulse is directed onto a grating and its first order diffraction is then focused by a cylindrical lens onto a mirror. The lens and mirror are mounted together on a platform which can be moved with respect to the grating with a micrometer linear stage, Fig.\ \ref{fig:pulse_shaper}(B). This allows a variation of the lens--grating separation while keeping the mirror in the focal plane of the lens. A lens--grating separation larger than the focal length (positive displacement) introduces a negative chirp, a separation smaller than the focal length (negative displacement) results in a positive chirp \cite{Martinez1987}. We demonstrate in Fig.\ \ref{fig:pulse_shaper}(B) a stretching of the original, transform-limited \SI{130}{\femto\second} pulses to a duration of more than \SI{10}{\pico\second}, an increase by a factor of 100.

The pulse duration is linked to the chirp. The electric field of the unchirped pulse can be described with a Gaussian envelope in the time domain:
\begin{align}
	E(t)=&E_{0} \exp{\left(\frac{-t^2}{2\tau_{0}^2}-i\omega_{0}t\right)} \nonumber \\
	\Delta t_{0,\mathrm{FWHM}}^I=&2 \sqrt{\ln{2}} \: \tau_{0} \nonumber
\end{align}
where $\Delta t_{0,\mathrm{FWHM}}^I$ is the full-width-at-half-maximum (FWHM) of the pulse intensity. $\Delta t_{0,\mathrm{FWHM}}^I=\SI{130}{\femto\second}$ here. Equivalently, the unchirped pulse can be described in the frequency domain:
\begin{align}
	E(\omega)=&E_{0} \tau^2 \exp{\left[-\frac{\tau_{0}^2}{2}(\omega-\omega_{0})^2 \right]} \nonumber \\
	\Delta \omega_{0,\mathrm{FWHM}}^I=& \frac{2\sqrt{\ln{2}}}{\tau_{0}}. \nonumber
\end{align}
The role of the pulse shaper is to add a phase for each frequency component:
\begin{align}
	\label{phase}
	\phi(\omega)=\phi_{0}+\phi_{1}(\omega-\omega_{0})+\frac{\phi_{2}}{2}(\omega-\omega_{0})^2.
\end{align}
The linear phase term shifts the entire pulse in time and is irrelevant here; the quadratic term represents a chirp. In practice, $\phi_2$ is directly proportional to the displacement of the pulse shaper \cite{weiner2009ultrafast}. The link between $\phi_2$ and the new pulse duration $\tau$ is:
\begin{subequations}
	\begin{align}
		\tau=&\sqrt{\tau_0^2+\left(\frac{\phi_2 }{\tau_{0}}\right)^2}\\
		\Delta t_{\mathrm{FWHM}}^I=&\sqrt{\left(\Delta t_{0,\mathrm{FWHM}}^I\right)^2+\left(\frac{\phi_2\ 4 \ln{2}}{\Delta t_{0,\mathrm{FWHM}}^I}\right)^2}.\label{eq:tau}
	\end{align}
\end{subequations}
A key parameter for adiabatic passage is the sweep rate $\alpha$. It is related to $\phi_2$ by
\begin{align}
		\frac{\mathrm{d}}{\mathrm{d} t} \omega(t)=\alpha=\frac{\phi_{2}}{\tau_{0}^4+\phi_{2}^2}.\label{eq:alpha}
\end{align}
We work here in the regime $\phi_{2} \gg \tau_{0}^{2}$ such that $\tau \simeq \phi_0/\tau_0$ and $\alpha \simeq 1/\phi_2$.

The measured pulse duration as a function of displacement is shown in Fig.\ \ref{fig:pulse_shaper}(B). The results can be described perfectly with a quadratic phase dependence, eq.\ \ref{phase}, allowing us to determine the sweep rate from eq.\ \ref{eq:alpha}. Our pulse shaper is capable of generating $\phi_{2}$ up to a few \SI{}{\pico\second\squared} with both signs. With a transform limited pulse duration of \SI{130}{\femto\second} and a chirp range \SIrange{0}{1}{\pico\second\squared} we can reach sweep rates from \SIrange{1}{80}{\per\pico\second\squared}. We note that the \SI{8}{\meter} long single mode fibers used to transport the laser pulses to the microscope in our experiment, Fig.\ \ref{fig:exp_scheme}, introduce a positive chirp of \SI{0.3\pm0.1}{\pico\second\squared}: this is compensated with our pulse shaper.

\subsubsection{Microscope design}\label{ssec:microscope}
\begin{figure*}[thbp]{}
	\centering
	\includegraphics[scale=1.0]{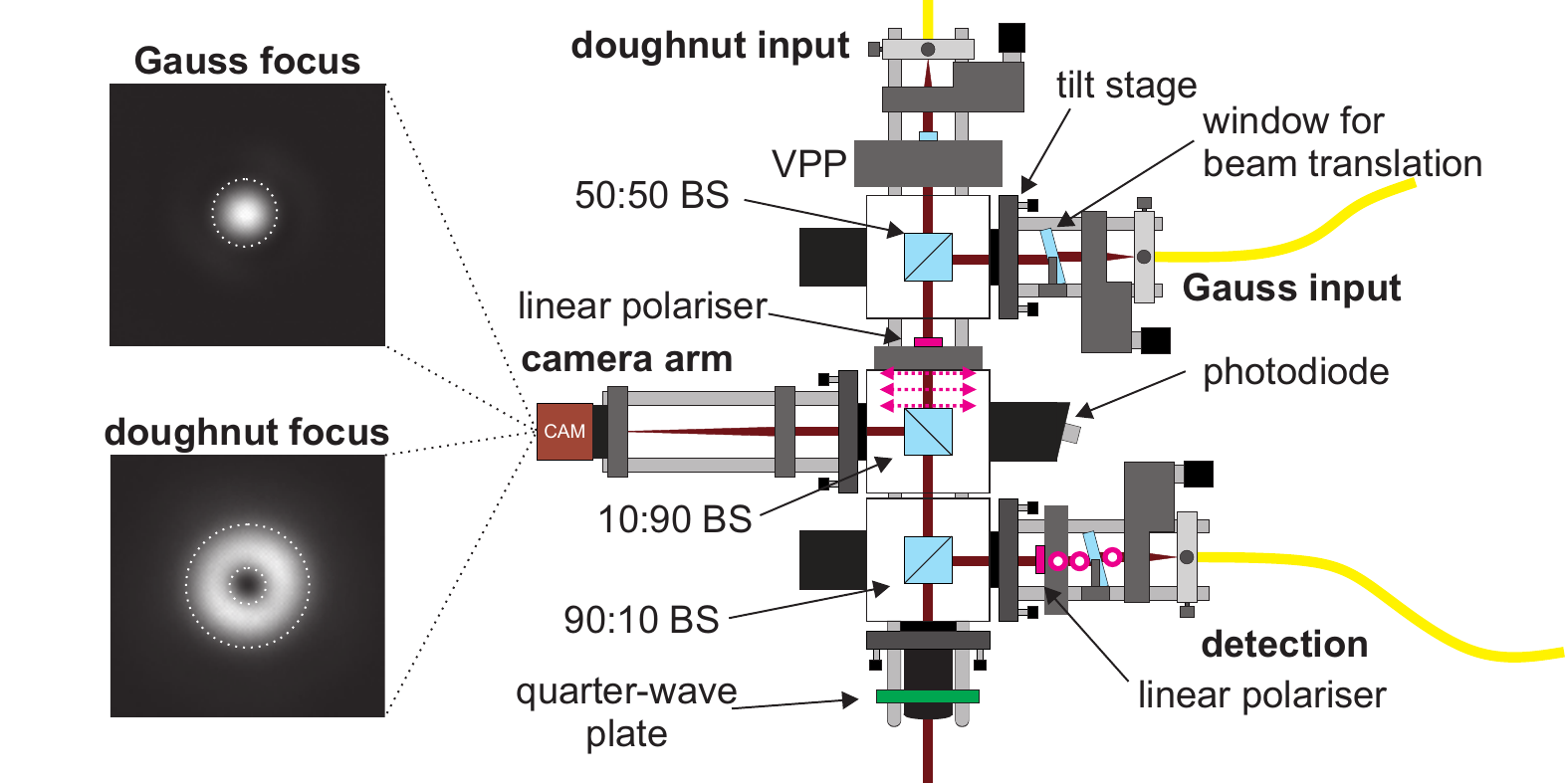}
	\caption{ \textbf{Microscope optics.} CAM camera, BS beam-splitter.
	}
	\label{fig:STED_microscope}
\end{figure*}
The microscope is based on a former version of an RF confocal microscope \cite{Kuhlmann2013_RSI}. The ``head", Fig.\ \ref{fig:STED_microscope}, accepts two fiber inputs. In each input, the exact propagation axis is controlled (by rotating a laser window and a tilt stage) and a linear polarisation state created (by a high quality polariser) after the two beams are combined by a beam-splitter and sent to the objective at low temperature. One input passes through a $2\pi$ vortex phase-plate (VPP). The VPP introduces a phase shift to the wavefront: the phase shift is equal to the in-plane radial angle. On focusing, a doughnut intensity profile results. The fiber output collects the back-scattered signal. A camera provides an in situ image: it assists in positioning the sample surface with respect to the focus, and in co-aligning the three beams (two inputs, one output). The objective is a single element aspheric lens with numerical aperture $\mathrm{NA_{O}}=0.68$ which, with CW laser input, can operate at close to the diffraction limit at these wavelengths.

Reflected laser light is suppressed by exciting and detecting in orthogonal polarisations, either in a linear polarisation basis as in previous experiments \cite{Kuhlmann2013_NatPhys,Kuhlmann2013_RSI}, or, as a new feature here, in a circular basis. We can make this choice by choosing the angle of a quarter wave-plate through which all beams pass, inserted as the final component in the microscope head. The extinction factor for the reflected laser light depends strongly on the quality of the focal spot. On a bare piece of GaAs we typically reach extinction ratios of \num{e8}:1; under a SIL on GaAs with CW excitation we reached in this experiment \num{e7}:1; and with excitation pulses, the single element aspheric lens, and a SIL on GaAs the ratio decreases to \num{e5}:1. The performance worsens with pulsed excitation because of the spectral bandwidth of the laser pulses: the aspheric lens introduces a slight chromatic aberration.

\begin{figure}[bhtp]{}
	\centering
	\includegraphics[scale=1.0]{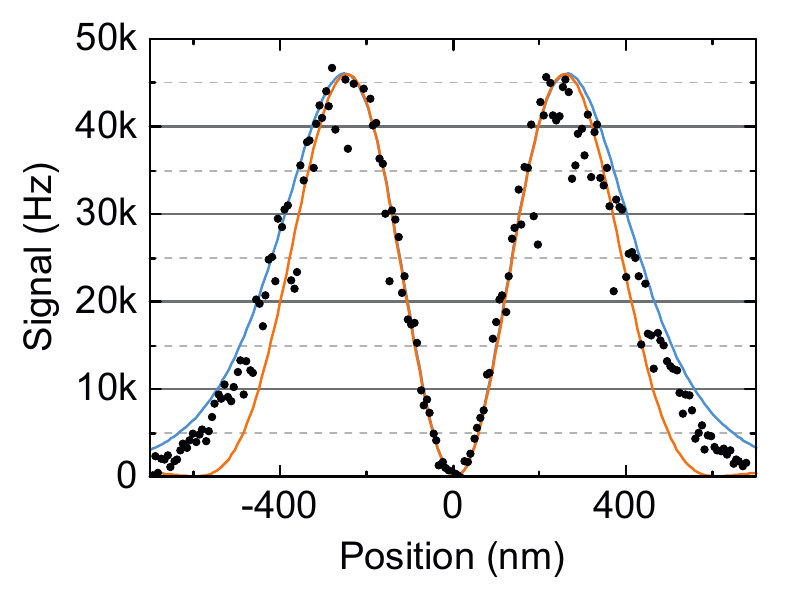}
	\caption{\textbf{Quality of the doughnut focus.} The doughnut-shaped focus was measured with a quantum dot and left circularly polarised excitation, data points (black circles) are taken from the white dashed line in Fig.\ \ref{fig:doughnut_displacement}. Also shown are calculated foci for a Gaussian (blue solid line) and plane wave (orange solid line) incident on the objective, $\left|E_{\mathrm{D}}\right|^2$ with $E_{\mathrm{D}}$ from eq.\ \ref{eq:doughnut}. Parameters for the calculation are $\lambda=\SI{940}{\nano\meter}$, $n_{\mathrm{SIL}}=2.13$, $\Delta X_{I,\mathrm{FWHM}}=\SI{2.0}{\milli\meter}$ ($\sigma=\SI{1.2}{\milli\meter}$), $f=\SI{5.9}{\milli\meter}$.
	}
	\label{fig:doughnut_quality}
\end{figure}

The quality of the microscope's focus is tested by imaging a chrome grid with confocal detection in reflection. With femto-second laser pulses with center wavelength $\lambda=\SI{940}{\nano\meter}$, we measured in this test experiment $\Delta x=\SI{388 \pm3}{\nano\meter}$ which is close to the diffraction limit.

The quality of the doughnut-shaped focus is crucial for the imaging experiment: a low quality reduces the intensity of the central bright spot. We define quality here as the intensity ratio of the doughnut's maximum to its minimum: it quantifies the relative residual intensity in the center,  Fig.\ \ref{fig:doughnut_quality}. Measured on a QD with left circularly polarised excitation we reached values of 1200:1, a factor of about five higher than in right circularly polarised excitation. The dependence of the quality on the helicity of the polarisation (with respect to the VPP) is in agreement with vector-field calculations \cite{Hao2010} and reported experimental results on fluorescent nanoparticles \cite{Neupane2013}. With the correct choice of the helicity the residual intensity at the center is very small in this experiment \cite{Neupane2013,Bokor2005}.

The microscope is simple to operate and robust: once aligned it remains aligned over several days.

\subsubsection{Resonance fluorescence detection: pulsed excitation}
\begin{figure*}[thbp]{}
	\centering
	\includegraphics[scale=1.0]{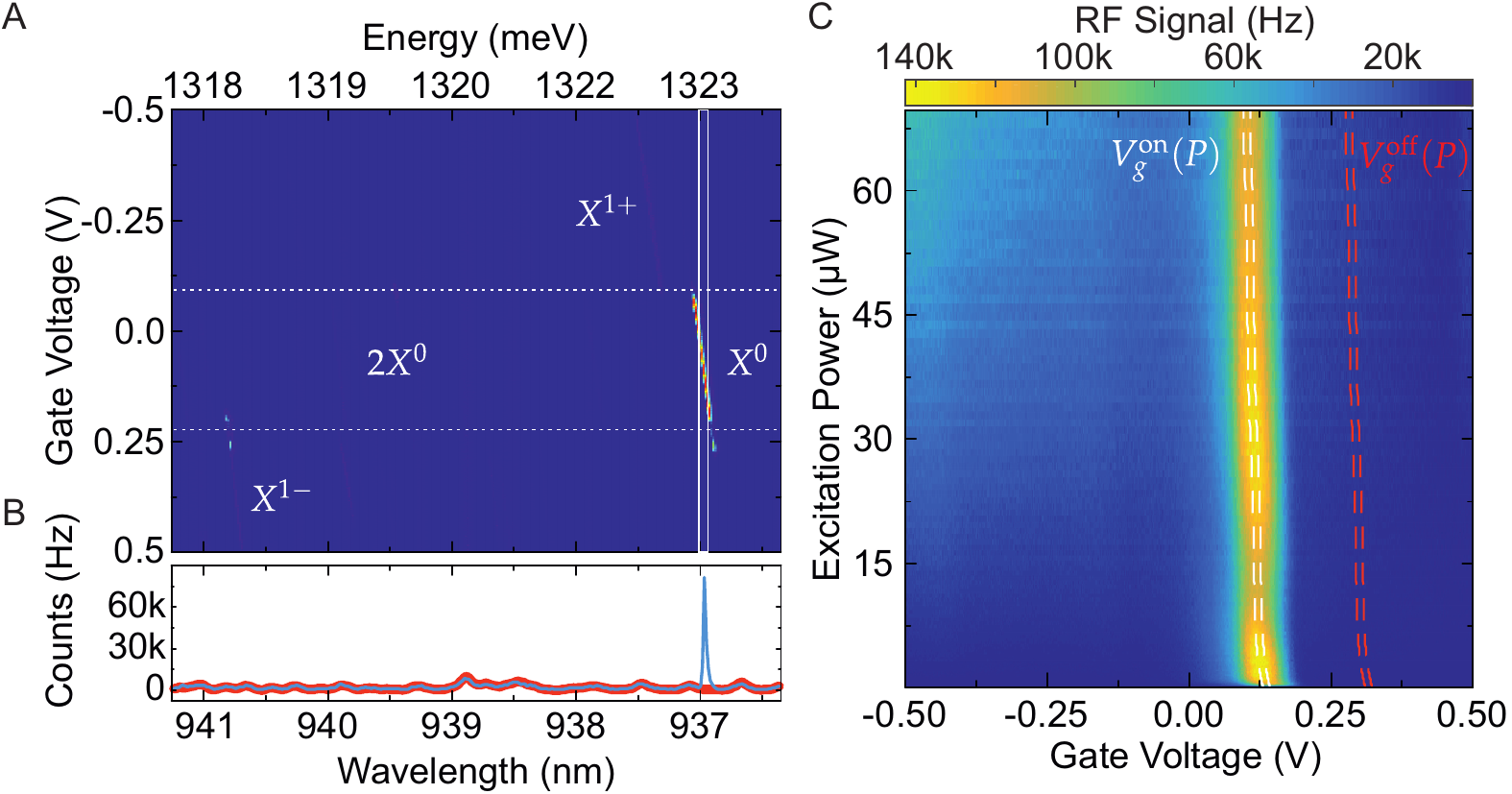}
	\caption{\textbf{Resonance fluorescence (RF) on a single quantum dot: pulsed excitation.} 
RF with positively chirped excitation pulses of a spectral and temporal width of $\Delta\lambda_{\mathrm{FWHM}}^I=\SI{9.70\pm0.30}{\nano\meter}$ and $\Delta t_{\mathrm{FWHM}}^I=\SI{7.0\pm0.3}{\pico\second}$, respectively. The polarisation was right circular, the integration time per spectrum \SI{25}{\milli\second}. 
	\textbf{(A)} Signal following background subtraction (see (B)) as a function of applied gate voltage $V_g$ and detection wavelength for an excitation power of \SI{0.36}{\micro\watt}. The detection of RF from $X^{1-}$, $X^{1+}$ and $2X^{0}$ after circularly polarised excitation is weak due to the cross polarised dark-field technique and the selection rules.
	\textbf{(B)} Signal versus wavelength: raw data at $V_{g}=\SI{120}{\milli\volt}$ (blue line). The laser background is determined from the whole dataset in (A): for every wavelength the $V_g$ with the smallest count rate is determined and this count rate is used to construct the laser background spectrum (red curve). Note that the laser suppression function is a quadratic function of wavelength about the alignment wavelength, superimposed with interference fringes.  
	\textbf{(C)} Spectrally integrated RF from the white solid rectangle in (A) as a function of excitation power $P$ and gate voltage. The final signal is the difference between the signal on the white dashed path, $V_g^{\mathrm{on}}(P)$, and the remaining background on the red dashed path, $V_g^{\mathrm{off}}(P)$.
	}
	\label{fig:pRF_PLE}
\end{figure*}

The femto-second pulses are spectrally broad: they contain the frequencies to excite not just the neutral exciton, X$^0$, but also the biexciton, 2X$^0$ and the trions, X$^{1-}$ and X$^{1+}$, should the ultra-short laser pulses disrupt the Coulomb blockade. These concerns turn out to be unfounded. With a circularly polarised pulsed excitation, the RF from a single quantum dot lies almost entirely at the X$^{0}$ emission frequency, Fig.\ \ref{fig:pRF_PLE}. The shift in emission frequency as a function of $V_g$ reflects the dc Stark effect.

For chirped excitation with sufficient intensity to invert the system, equivalently for a $\pi$ pulse with unchirped excitation, the RF:background ratio is $11:1$. At higher laser power, the background rises but the signal does not and the background signal dominates. We deal with this in two ways. First, the laser background is spectrally broad, the RF spectrally narrow, 5,000-times narrower in fact. We therefore collect the signal in a small spectral ``window", typically a sub-nanometer wavelength extent. We do this simply by reading out the appropriate pixels on the CCD camera. In this way, only the laser background in a narrow bandwidth adds to the signal. Secondly, we measure the remaining laser background by recording the signal when the quantum dot is turned off. We implement this by recording a signal at two different values of $V_g$: $V_g^{\mathrm{on}}$ and $V_g^{\mathrm{off}}$. This is a simple modulation technique and is robust with respect to any changes in the performance of the microscope. The only complication in this procedure is a slight shift of $V_g^{\mathrm{on}}$ as a function of excitation power: this arises because a small amount of space charge is trapped in the device and shifts the X$^{0}$ resonance. Since the excitation power is constant during a scan in the imaging experiment, here the simple modulation technique is sufficient. Other experiments (main paper, Fig.\ 2(C)) are performed as a function of pulse area. We therefore record here the signal-plus-background in a small gate voltage range around the X$^0$ resonance. The signal is then extracted from a path following the slight shift of the resonance, the background is determined from the same dataset and the same path but shifted in $V_g$ with respect to the resonance, Fig.\ \ref{fig:pRF_PLE}(C). This procedure enables us to determine the RF signal even at very large laser powers, corresponding to a pulse area up to $11 \pi$. The laser background contributes additional shot noise to the noise in the RF signal; there is also a systematic error which we estimate to be 20\% of the signal at the highest laser powers, arising from weak interference effects of the two pulses in the imaging experiment and from a slight dependence in the choice of X$^0$ spectral bandwidth.

\subsection{Imaging of two-level system with adiabatic passage: model}
We construct a simple model of nanoscopic imaging of a quantum mechanical two-level system. The aim is to capture the essential features and to generate analytical results. 

\subsubsection{The optical fields: scalar theory}
We calculate the field distribution in the focal plane using scalar diffraction theory applied to Gaussian beams. Strictly speaking, at high numerical apertures a full vector theory is required; additionally the beams are truncated Gaussians, truncated by the aperture of the objective lens. Extending the calculation to arbitrary input beams with a full vector theory is not difficult. The present approach has the virtue of simplicity.

\begin{figure}[thbp]{}
	\centering
	\includegraphics[scale=1.0]{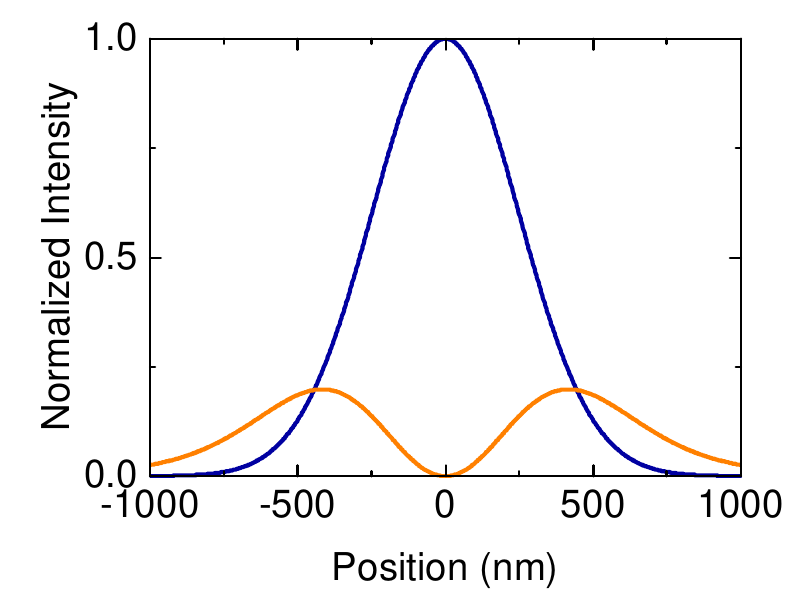}
  \caption{{\bf Gauss and doughnut intensity distribution.} $\left|E_{\mathrm{G}}(x,y=0)\right|^2$ and $\left|E_{\mathrm{D}}(x,y=0)\right|^2$ in eq.\ \ref{eq:fields} are plotted with $E_{\mathrm{G}}^{0}=E_{\mathrm{D}}^{0}=1$ in blue and orange, respectively. Parameters for the calculation are $\lambda=\SI{940}{\nano\meter}$, $n_{\mathrm{SIL}}=2.13$, $\Delta X_{I,\mathrm{FWHM}}=\SI{2.0}{\milli\meter}$ ($\sigma=\SI{1.2}{\milli\meter}$), $f=\SI{5.9}{\milli\meter}$.
 	}
  \label{fig:Gauss_doughnut_1d}
\end{figure}

In scalar theory, the electric field in the focal plane has the same polarisation as the input polarisation. The spatial distribution of the electric field in the focal plane is proportional to the Fourier transform of the (normalized) aperture function, $A(R,\phi)$, where $(R,\phi)$ are the radial coordinates in the plane of the aperture. For the Gaussian and doughnut beams
\begin{align}
A_{\mathrm{G}}(R,\phi)=&\frac{1}{2 \pi \sigma^2} \exp \left(- \frac{R^2}{2 \sigma^2} \right) \nonumber \\
A_{\mathrm{D}}(R,\phi)=& A_{\mathrm{G}}(R,\phi) \, e^{i \phi} \nonumber
\end{align}
where $\sigma$ describes the spatial extent of the beams before the focusing, $\Delta X_{I,\mathrm{FWHM}}=2\sqrt{\ln 2}\ \sigma$. The fields in the focal plane $(x,y)$ are given by
\begin{align}
	E_{\mathrm{G}}(r)&= E_{\mathrm{G}}^{0} \int_{R=0}^{\infty}\!\int_{\phi=0}^{2\pi} \! A_{\mathrm{G}} \exp{\left(-  ik\frac{r}{f} R \cos\phi\right)} R \, \mathrm{d}R \, \mathrm{d}\phi \nonumber \\
	E_{\mathrm{D}}(r)&= E_{\mathrm{D}}^{0} \int_{R=0}^{\infty}\!\int_{\phi=0}^{2\pi} \! A_{\mathrm{D}} \exp{\left(-ik \frac{r}{f} R \cos\phi\right)} R \, \mathrm{d}R \, \mathrm{d}\phi \nonumber
\end{align}
where $r=\sqrt{x^2+y^2}$ is the radial coordinate in the focal plane, $k=2 \pi /(\lambda/n_{\mathrm{SIL}})$, $f$ is the focal length of the lens, and $E_{\mathrm{G}}^{0}$ ($E_{\mathrm{D}}^{0}$) is the amplitude of the Gauss (doughnut) beam.
\begin{subequations}
\label{eq:fields}
	\begin{align}
		E_{\mathrm{G}}(r)&=E_{\mathrm{G}}^{0} \, \exp{\left(-\frac{k^2 \sigma^2 r^2}{2 f^2}\right)}\label{eq:GDbeams}\\
		E_{\mathrm{D}}(r)&=E_{\mathrm{D}}^{0} i \frac{\sqrt{\pi}}{2 \sqrt{2}} \frac{k \sigma r}{f}  \exp{\left(-\frac{k^2 \sigma^2 r^2}{4 f^2}\right)}\nonumber \\
		&\cdot \left[I_0\left( \frac{k^2 r^2 \sigma^2}{4 f^2}\right)-I_1\left( \frac{k^2 r^2 \sigma^2}{4 f^2}\right)\right].\label{eq:doughnut}
	\end{align}
\end{subequations}
$I_0$ and $I_1$ are Bessel functions of the first kind with order 0 and 1, respectively. $E_{\mathrm{G}}(x,y)$ and $E_{\mathrm{D}}(x,y)$ are functions of $(x,y)$ and have a Gaussian and a doughnut shape, respectively. The corresponding intensity distributions are depicted in Fig.\ \ref{fig:Gauss_doughnut_1d}. The FWHM of the Gaussian intensity distribution is $\Delta x_{I,\mathrm{FWHM}}=2\sqrt{\ln 2}\ f/(k \sigma)$. $E_{\mathrm{D}}(r)$ is zero at $r=0$.

\subsubsection{Adiabatic passage: the Landau--Zener model}
\begin{figure*}[thbp]{}
	\centering
	\includegraphics[scale=1.0]{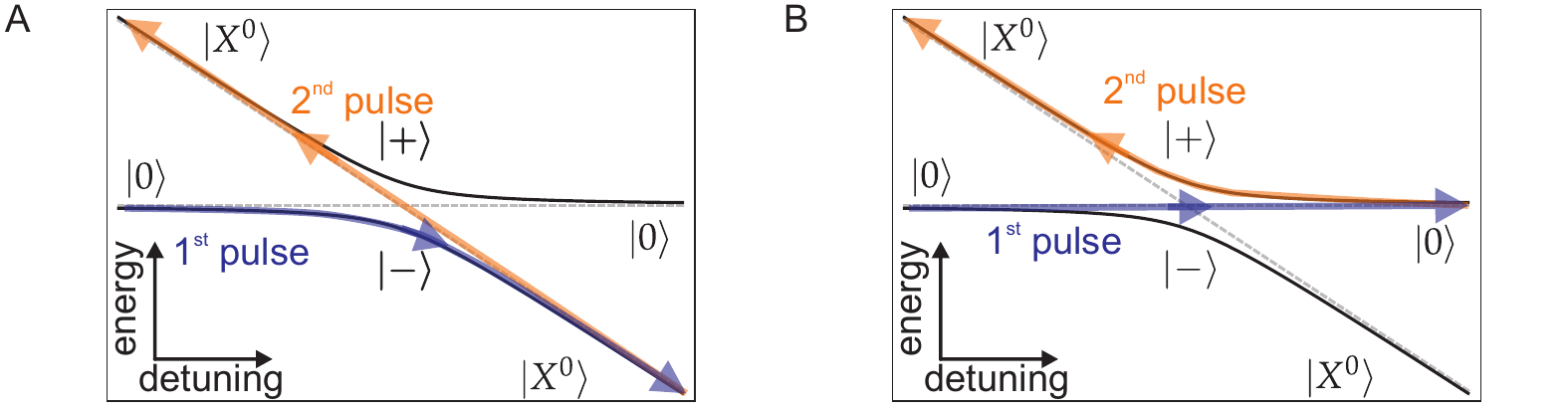}
  \caption{\textbf{Two-pulse adiabatic passage on a two-level system.} The eigenenergies are plotted as a function of detuning. Two possible paths, (A) and (B), show the two routes by which the system starts in the lower state $\ket{0}$ and ends up in the upper state $\ket{1}$ after a two pulse sequence. \textbf{(A)} Adiabatic passage followed by tunnelling; \textbf{(B)} tunnelling followed by adiabatic passage.
 	}
  \label{fig:STED_sim_paths}
\end{figure*}

The case of a driven two-level system with time-dependent detuning was considered a long time ago by Landau \cite{Landau} and by Zener \cite{Zener1932}. Initially, the system is in one of the basis states, $\ket{0}$ or $\ket{1}$, and the detuning of the drive is large and negative. The detuning is then increased at a fixed rate passing through zero until it is large and positive. As a function of detuning, there is an avoided crossing in the eigenvalues, Fig.\ \ref{fig:STED_sim_paths}. The probability that the system makes a transition from one basis state to the other is given by the Landau--Zener factor, $p_{\mathrm{LZ}}$:
\begin{equation}
	p_{\mathrm{LZ}}(\Omega,\alpha)=\exp{\left(-\frac{\pi \Omega^2}{2\left|\alpha\right|}\right)} \nonumber
\end{equation}
where $\Omega$ is the angular Rabi frequency, $\alpha$ the sweep rate (rate of change of the angular frequency). Here, the coupling arises from the interaction of the optical dipole $\mu_{01}$ with the electric field of the light wave, $\hbar \Omega = \mu_{01} E$. In the ``sudden" regime when $p_{\mathrm{LZ}} \simeq 1$, the system ``tunnels" through the avoided crossing and $\ket{0} \rightarrow \ket{0}$, $\ket{1} \rightarrow \ket{1}$. Alternatively, in the limit when $p_{\mathrm{LZ}} \ll 1$, the states are swapped $\ket{0} \rightarrow \ket{1}$, $\ket{1} \rightarrow \ket{0}$: this is adiabatic passage, Fig.\ \ref{fig:STED_sim_paths}. The imaging protocol exploits the exponential dependence of $p_{\mathrm{LZ}}(\Omega,\alpha)$ on $\Omega$ for fixed $\alpha$: there is a fast transition from the sudden regime to the adiabatic passage regime as $\Omega$ increases.

In our case here, the excitation is not just chirped but also pulsed. In other words, $\Omega$ is a function of time which is not a feature of the Landau--Zener model. However, the Landau--Zener model captures the essential physics. The Rabi energies are at most a few meV in these experiments yet the spectral bandwidth of the pulse is $\sim 15$ meV. The peak of the laser spectrum is chosen to lie close to the X$^0$ transition. This means that in the region of the avoided crossing, the Rabi frequency is approximately constant in time, such that the Landau--Zener model is a reasonable one to describe the experiment. Of course, it is possible to integrate the Schr\"{o}dinger equation over the pulse \cite{Simon2011}. The advantage of the Landau--Zener model is that there is a simple analytical result.

The imaging experiment depends on the response to two pulses, the Gauss-pulse (Rabi coupling $\Omega_{\mathrm{G}}$, chirp $\alpha_{\mathrm{G}}$) and the doughnut-pulse (Rabi coupling $\Omega_{\mathrm{D}}$, chirp $\alpha_{\mathrm{D}}$). $\Omega_{\mathrm{G}}$ and $\Omega_{\mathrm{D}}$ depend on the lateral coordinates $(x,y)$ in the focal plane: this enables imaging. In practice, $\alpha_{\mathrm{G}}>0$, $\alpha_{\mathrm{D}}=-\alpha_{\mathrm{G}}$. The system is initially in the ground state, $\ket{0}$. There are two possible routes for the system to end up in the excited state, $\ket{1}$, Fig.\ \ref{fig:STED_sim_paths}. (In practice in the experiment, $\ket{1} \equiv \ket{\mathrm{X} ^{0}}$.) Adding together the probabilities, the probability of occupying state $\ket{1}$ with the two-pulse combination is
\begin{eqnarray*}
		P_1(\Omega_{\mathrm{G}},\alpha_{\mathrm{G}},\Omega_{\mathrm{D}},\alpha_{\mathrm{D}})=&p_{\mathrm{LZ}}(\Omega_{\mathrm{G}},\alpha_{\mathrm{G}})+p_{\mathrm{LZ}}(\Omega_{\mathrm{D}},\alpha_{\mathrm{D}}) \\
		&-2p_{\mathrm{LZ}}(\Omega_{\mathrm{G}},\alpha_{\mathrm{G}})\,p_{\mathrm{LZ}}(\Omega_{\mathrm{D}},\alpha_{\mathrm{D}}). 
\end{eqnarray*}
The probability that the two-level system emits a photon following the two-pulse combination is $P_1$. The probability that the photon is detected is the point-spread-function of the detection channel, a Gaussian function centerd at $(x,y)=(0,0)$ (the square of eq.\ \ref{eq:GDbeams}), multiplied by a constant factor $\beta$, the efficiency of the entire microscope--detector system. We arrive at an analytical result for the imaging signal: 
\begin{align}
	\mathrm{RF}(x,y)=&\beta \, P_1(\Omega_{\mathrm{G}},\alpha_{\mathrm{G}},\Omega_{\mathrm{D}},\alpha_{\mathrm{D}}) \nonumber \\ 
									 & \cdot \exp{\left(-\frac{k^2 \sigma^2 (x^2+y^2)}{f^2}\right)} \label{eq:calculatedRF}
\end{align}

\begin{figure}[bhtp]{}
	\centering
	\includegraphics[scale=1.0]{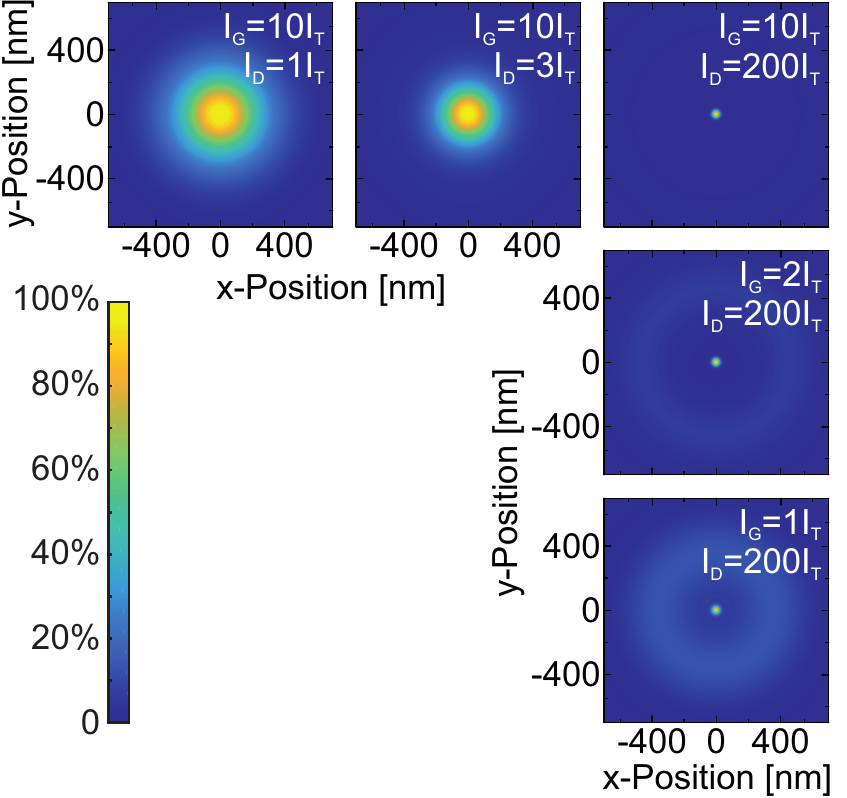}
  \caption{{\bf Calculated images.} The function ${\mathrm{RF}}(x,y)$, eq.\ \ref{eq:calculatedRF}, is plotted for $\lambda=\SI{940}{\nano\meter}$, $n_{\mathrm{SIL}}=2.13$, $\Delta X_{I,\mathrm{FWHM}}=\SI{2.0}{\milli\meter}$ ($\sigma=\SI{1.2}{\milli\meter}$), $f=\SI{5.9}{\milli\meter}$, $\alpha_{\mathrm{G}}=-\alpha_{\mathrm{D}}=\SI{3.24}{\pico\second^{-2}}$. For large Gauss beam intensity, the central spot decreases in size retaining the maximum signal as the doughnut beam intensity increases. For large doughnut beam intensity, a ring surrounds the central spot at small Gauss beam intensity. The ring weakens and moves to larger radii as the Gauss beam intensity increases. 
 	}
  \label{fig:STED_sim_images}
\end{figure}
The $\mathrm{RF}$ function is plotted in Fig.\ \ref{fig:STED_sim_images} as a function of $(x,y)$ for three different values of $\Omega_{\mathrm{G}}^0$ and three different values of $\Omega_{\mathrm{D}}^0$. For large $\Omega_{\mathrm{G}}^0$, a bright spot appears at $(x,y)=(0,0)$ whose width decreases as $\Omega_{\mathrm{D}}^0$ increases: this describes the main modus operandi of the imaging scheme. At intermediate $\Omega_{\mathrm{G}}^0$, a ring appears surrounding the central maximum. The ring appears at locations where the Gauss beam leaves part of the population in state $\ket{0}$, and the doughnut beam transfers part of this population into state $\ket{1}$, resulting in a weak residual signal. As $\Omega_{\mathrm{G}}^0$ increases the ring moves to larger radii in the focal plane and it weakens: it can be suppressed in practice simply by choosing a large enough Gauss beam intensity.

In the limit of large $\Omega_{\mathrm{G}}^0$ and $\Omega_{\mathrm{D}}^0$, it is possible to derive a result for the FWHM of the central maximum, $\Delta x_{\mathrm{FWHM}}$. We assume that $\Omega_{\mathrm{G}}^0$ is large enough such that $p_{\mathrm{LZ}}(\Omega_{\mathrm{G}}) \simeq 1$ over the central bright spot. Expanding the doughnut field $E_{\mathrm{D}}(x,y)$ as a Taylor series in $(x,y)$ and retaining only the linear term, we find
\begin{equation}
	\Delta x_{\mathrm{FWHM}}=\frac{\Delta x_{I,\mathrm{FWHM}}^0}{\sqrt{1+\pi^2 (\Omega_{\mathrm{D}}^0)^2/16 \alpha}} \nonumber
\end{equation}
This has the same dependence on the doughnut beam intensity as the STED protocol \cite{Harke2008}.

\begin{figure}[htbp]{}
	\centering
	\includegraphics[scale=1.0]{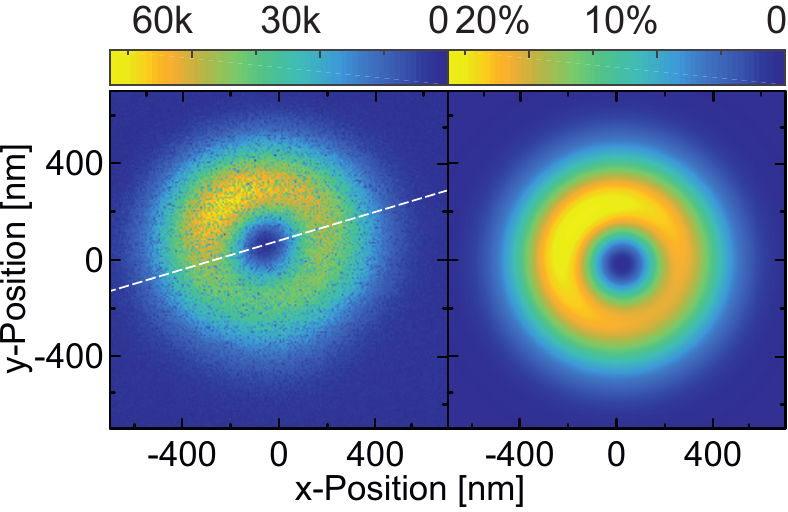}
  \caption{{\bf Doughnut beam displacement.} Left: Resonance fluorescence from quantum dot QD01 measured with the doughnut-shaped pulse alone and left circularly polarised excitation. The excitation intensity was $I_{D}^0=1.2I_T$. A cut along the white dashed line is plotted in Fig.\ \ref{fig:doughnut_quality}. Right: corresponding simulation with $I_{D}^0=1.2I_T$, parameters as stated in Fig.\ \ref{fig:STED_sim_images} and a displacement of $\Delta x_{D}=\SI{20}{\nano\meter}$ and $\Delta y_{D}=\SI{-20}{\nano\meter}$.}
  \label{fig:doughnut_displacement}
\end{figure}

As a final point, we note that the response to the doughnut beam alone can exhibit a radial asymmetry, Fig.\ \ref{fig:doughnut_displacement}. This arises as a consequence of a small displacement of the doughnut zero with respect to the collection point-spread-function. By introducing a small displacement in the model, we can reproduce the experimental results convincingly, Fig.\ \ref{fig:doughnut_displacement}. Likewise, the ring in the calculated ${\mathrm{RF}}(x,y)$ function for intermediate $\Omega_{\mathrm{G}}^0$ also becomes radially asymmetric in the presence of this displacement. However, in the main imaging regime, the central bright spot lies at the central zero of the doughnut beam and is only very weakly influenced by a lateral displacement of the Gauss beam, equivalently the collection point-spread-function. The main experimental requirement is therefore to create a high quality doughnut beam. 

\end{document}